\documentclass{nature}

\usepackage{graphicx}
\title{Spatially Resolved Magnetic Field Structure in the Disk of a T Tauri Star}

\author{Ian W. Stephens$^{1,2}$, Leslie W. Looney$^2$, Woojin Kwon$^3$, Manuel Fern\'{a}ndez-L\'{o}pez$^{2,4}$, A. Meredith Hughes$^5$, Lee G. Mundy$^6$, Richard M. Crutcher$^2$, Zhi-Yun Li$^7$, \& Ramprasad Rao$^8$}

\begin{document}

\newcommand\arcsec{\mbox{$^{\prime\prime}$}}%
\maketitle

\begin{affiliations}
 \item Institute for Astrophysical Research, Boston University, Boston, MA 02215, USA
 \item Department of Astronomy, University of Illinois, Urbana, Illinois 61801, USA
 \item SRON Netherlands Institute for Space Research, Landleven 12, 9747 AD Groningen, The Netherlands
 \item Instituto Argentino de Radioastronom{\'i}a, CCT-La Plata (CONICET), C.C.5, 1894, Villa Elisa, Argentina
 %\item Van Vleck Observatory, Astronomy Department, Wesleyan University, 96 Foss Hill Drive, Middletown, CT 06459, USA
 \item Van Vleck Observatory, Astronomy Department, Wesleyan University, Middletown, CT 06459, USA
 \item Astronomy Department \& Laboratory for Millimeter-wave Astronomy, University of Maryland, College Park, MD 20742, USA
 \item Astronomy Department, University of Virginia, Charlottesville, VA 22904, USA
 %\item Institute of Astronomy and Astrophysics, Academia Sinica, 645 N. Aohoku Place, Hilo, HI 96720, USA
 \item Institute of Astronomy and Astrophysics, Academia Sinica, Hilo, HI 96720, USA
\end{affiliations}
\noindent \emph{Accepted for publication in Nature}
\begin{abstract}

Magnetic fields in accretion disks play a dominant role during the star formation process\cite{Blandford1982,Balbus1998} but have hitherto been observationally poorly constrained. Field strengths have been inferred on T Tauri stars themselves\cite{JK2007} and possibly in the innermost part of the accretion disk\cite{Donati2005}, but the strength and morphology of the field in the bulk of the disk have not been observed. Unresolved measurements of polarized emission (arising from elongated dust grains aligned perpendicular to the field\cite{Lazarian2007}) imply average fields aligned with the disks\cite{Tamura1995,Tamura1999}. Theoretically, the fields are expected to be largely toroidal, poloidal, or a mixture of the two\cite{Balbus1998,Cho2007,Blandford1982,Konigl2000,Hennebelle2009}, which imply different mechanisms for transporting angular momentum in the disks of actively accreting young stars such as HL Tau\cite{Robitaille2007}. Here we report resolved measurements of the polarized 1.25 mm continuum emission from HL Tau's disk. The magnetic field on a scale of 80 AU is coincident with the major axis ($\sim$210 AU diameter\cite{Kwon2011}) of the disk. From this we conclude that the magnetic field inside the disk at this scale cannot be dominated by a vertical component, though a purely toroidal field does not fit the data well either. The unexpected morphology suggests that the magnetic field's role for the accretion of a T Tauri star is more complex than the current theoretical understanding.

%HL Tau is a young T Tauri star with strong accretion onto the star\cite{Robitaille2007} which is likely driven by magnetohydrodynamical turbulence\cite{Balbus1998}.
\end{abstract}

HL Tau is located 140 pc away\cite{Rebull2004} in the Taurus molecular cloud. Although HL Tau is a T~Tauri star, it is considered to be an early example due to its bipolar outflow\cite{Movsessian2012} and possible residual envelope\cite{Welch2000}. Observations and modeling of the protostar with a thick, flared disk suggest a stellar mass of $\sim$0.55~$M_\odot$ and a disk mass of 0.14~$M_\odot$\cite{Kwon2011}. A possible planet forming in HL Tau's disk has been observed\cite{Greaves2008}, though this detection was not confirmed\cite{Kwon2011}. However, HL Tau's disk is gravitationally unstable which could favor fragmentation into Jupiter mass planets\cite{Kwon2011,Greaves2008}. HL Tau has the brightest T Tauri star disk at millimeter wavelengths, allowing for the best possible probe of the fractional polarization $P$. Previous observations of the polarization of HL Tau's disk yielded marginally significant, spatially unresolved polarization detections with the James Clerk Maxwell Telescope (JCMT, $P$ = 3.6\% $\pm$ 2.4\% at 14$\arcsec$ = 1960 AU resolution)\cite{Tamura1995} and the Submillimeter Array (SMA, $P$ = 0.86\% $\pm$ 0.4\% at 2$\arcsec$ = 280 AU resolution, archival observations released in this paper). In addition, HL Tau observations with the Combined Array for Millimeter-wave Astronomy (CARMA) have shown that the interferometric emission comes entirely from the disk with no contamination of large-scale envelope emission\cite{Kwon2011}. HL Tau is therefore a very promising source to search for a resolved polarization detection.

% and is at the age at which planets should start to form. 

%Only through observations of polarized dust emission can the morphology of the magnetic field be ascertained. Since the long axis of dust grains preferentially align perpendicular to the magnetic field, the PA of the dust polarization is orthogonal to the PA of the magnetic field, $\theta_B$\cite{Lazarian2007}.

%Though unresolved observations (with resolutions 5-10 times larger than the disk) toward TTSs suggest toroidal fields with the fraction of polarized light, $P$, are found to be $\sim$2-3\%\cite{Tamura1995,Tamura1999}, polarimetric measurements that spatially resolve disks around TTSs and Herbig AE stars have placed stringent upper limits ($P<1$\%) on the polarization fraction of dust continuum emission\cite{Hughes2009,Hughes2013}, which disagree with theoretical models of high efficiency grain alignment with a purely toroidal field\cite{Cho2007}. 

Only through observations of polarized dust emission can the morphology of the magnetic field be ascertained, though higher resolution dust polarimetric observations of T Tauri star disks do not detect polarization and place stringent upper limits ($P<1$\%) on the polarization fraction\cite{Hughes2009,Hughes2013}, which disagree with theoretical models of high efficiency grain alignment with a purely toroidal field\cite{Cho2007}. There is a clear discrepancy between theoretical models of the magnetic fields in disks and the observations to-date, requiring more sensitive observations of the dust polarization. The SMA recently detected the magnetic field morphology in the circumstellar disk of the Class 0 (i.e., the earliest protostellar stage) protostar IRAS~16293--2422~B\cite{Rao2014}, but since this disk is nearly face-on, observations cannot detect the vertical component of the magnetic field. Moreover, this source is perhaps one of the youngest of the known Class 0 sources\cite{Loinard2013}, increasing the chances that the polarized flux could be from the natal environment. Nevertheless, since the disk is the brightest component at the probed scale, polarization most likely comes from the disk, and the magnetic field morphology hints at toroidal wrapping\cite{Rao2014}. 

%HL Tau has had marginal, unresolved polarization detections with the James Clerk Maxwell Telescope (JCMT, $P=3.6\pm2.4\%$ at 14$\arcsec$ (1960 AU) resolution)\cite{Tamura1995} and the SMA ($P=0.86\pm0.4\%$ at 2$\arcsec$ (280 AU) resolution). Additionally, HL Tau is the brightest TTS disk at millimeter wavelengths, allowing for the deepest possible probe on $P$. HL Tau also has strong accretion\cite{Robitaille2007} and an infalling envelope\cite{DAlessio1997}, and since accretion is considered to be associated with MHD turbulence, such processes may aid in a polarization signal. Therefore, HL Tau is a quintessential source to search for a polarization detection for TTSs.

%Interferometric observations of HL Tau are also insensitive to large-scale structure, ensuring that polarized flux comes entirely from the dis
%HL Tau also exhibits strong accretion\cite{Robitaille2007} and an infalling envelope\cite{DAlessio1997}, both of which are typically associated with MHD turbulence that is likely to produce a polarization signal. 

%high-resolution observations of three TTS (TW Hydrae, DG Tau, and GM Tau) and two Herbig AE stars (HD 163296 and MWC 480) 

%The CARMA dual polarization receivers allow for the measurement of polarized dust emission. Polarimetric maps can thus be created, which provide the flux density (Stokes $I$), the position angle (PA, measured counterclockwise from north) of the dust polarization, and $P$ at every point within the map. 

We have obtained 1.25~mm CARMA polarimetric maps of HL Tau at 0.6$\arcsec$ (84 AU) resolution and plotted the magnetic field morphology in Figure 1. This is a resolved detection (with approximately 3 independent beams) of the magnetic field morphology in the circumstellar disk of a T Tauri star. The central magnetic field vector has a measured position angle (PA, measured counterclockwise from north) of $\theta_B = 143.6^\circ \pm 4.4^\circ$, which is within 9$^\circ$ of the previously measured PA of the major axis of the HL Tau disk\cite{Kwon2011}. This angle is in agreement with that measured by the JCMT ($\theta_B = 140^\circ \pm 20^\circ$)\cite{Tamura1995} and archival observations from the SMA analyzed here ($\theta_B = 137^\circ \pm 13^\circ$). The central vector has a fractional polarization of $P=0.59\% \pm0.09\%$, and $P$ varies over the disk between $0.54\%\pm0.13$\% and $2.4\%\pm0.7$\% with an average and median $P$ of 0.90\% and 0.72\% respectively, in agreement with the upper limits ($P<1$\%) of other T Tauri star polarimetric observations\cite{Hughes2009,Hughes2013}. This median value is significantly less than IRAS~16293--2422~B (1.4\%) which could indicate disk evolution; the dust grains could grow larger and become more spherical with time, and/or the magnetic field is becoming more turbulent.

%which could indicate that the disk is forming larger and more spherical dust grains and/or that the magnetic field is more tangled.

%These high resolution interferometric observations are insensitive to large scale structure. Single dish and interferometric HL Tau observations find very similar compact fluxes, suggesting that the envelope dust continuum is negligible\cite{Lay1997}, and the flux appears to come entirely from the disk\cite{Kwon2011}. Therefore, it is extremely likely that the polarization comes from the disk.

%Unresolved observations of a toroidal magnetic field in a disk will measure an averaged field aligned with the major axis; however, once the disk is resolved, a toroidal field will have more structure.

To constrain the intrinsic magnetic field configuration inside HL Tau's disk, we made a simple model which incorporates a combination of a toroidal and a vertical component; a radial field component inside the bulk of the disk (most likely probed by our polarization observation) is expected to be sheared quickly into a toroidal configuration by differential rotation on the short time-scale of the disk rotation. We use the best estimates of disk parameters\cite{Kwon2011} (PA = $136^\circ$ and inclination $i=40^\circ$, both accurate within a few degrees) and vary the relative strength of the two field components from 100\% toroidal and 0\% vertical to 0\% toroidal and 100\% vertical in 1\% increments (modeling details are in the Methods section). Using the $\geq$3$\sigma$ detections in Figure 1, we find that a completely toroidal field is the best fit model.  However, the reduced $\chi^2$ value is high (69, where a value of 1 indicates an optimal fit). If we do not constrain the magnetic field to be aligned with the best-fit disk\cite{Kwon2011} we can achieve a better fit. For a disk with a PA of $151^\circ$, the best fit model to the observations is almost completely edge on, with an inclination of $i=89^\circ$ (reduced $\chi^2$ value of 1). Figure 2 shows the observed versus the modeled parameters for PA of $136^\circ$ and $i=40^\circ$ as well as PA of $151^\circ$ and $i=89^\circ$. While the PA does not vary largely from the major axis of the disk ($15^\circ$ difference), $i=89^\circ$ is inconsistent with our disk dimensions and the well-constrained continuum observations of other studies\cite{Lay1997,Kwon2011}. There is also no detected polarized flux in the northeast (upper-left) and southwest (bottom-right) of the disk where $P$ for a toroidal field should be the highest due to less beam smearing. This fact, along with the poor fit with the constrained disk parameters, indicates that although there is probably a toroidal component, the field apparently has substantial contributions from other components.

%In order to determine the contribution of the toroidal and poloidal field components in HL Tau, we modeled a mix of the two components.
%Starting from a modeled field of 100\% toroidal and 0\% poloidal, we decremented the toroidal field by 1\% and incremented the poloidal field by 1\% (99\% toroidal and 1\% poloidal, 98\% toroidal and 1\% poloidal, ...) all the way to 0\% toroidal and 100\% poloidal for a total of 101 modeled maps.

We note that adding a vertical component to all our models makes the fit worse. The lack of any vertical component in the model suggests that dominant poloidal fields are likely absent inside the disk since, as we argued earlier on dynamical grounds, a predominantly radial field is unlikely inside the disk because of rapid differential rotation. Moreover, although points with $>$5$\sigma$ polarization detections fit a toroidal morphology even for the $i=40^\circ$ model, we cannot be sure that the disk field is only toroidal since a straight, uniform field in the plane of the disk, although physically unlikely for a disk system, also well fits the data. %Nevertheless, there could be a toroidal component and likely rule out the possibility of poloidal fields at the 80~AU scale (i.e., poloidal fields may be present at smaller scales). Theory has suggested the scenario that magnetic fields threading the disk are wrapped in a toroidal configuration by differential rotation of the envelope and/or disk (although such a scenario does not necessarily preclude poloidal fields\cite{Hennebelle2009}). %However, these observations by no means completely confirm toroidal fields due to the poor fits with the best model.

%What causes the outer vectors to not fit the toroidal morphology based on HL Tau's best fit disk dust model\cite{Kwon2011} (Figure 1, left)? This question is difficult to answer with these observations, though we can speculate. Perhaps the inner disk, where the outflow originates, is warped compared to an edge-on outer disk. In order to fit a warp to the high-resolution 1~mm continuum observations\cite{Kwon2011}, the outer disk has to be a fat toroid that is much thicker than the much smaller inner disk. Large warps have been suggested by both theory and observations\cite{brinch2007,facchini2013}. Another possibility is that the magnetic field in the outer part of the disk is influenced by the magnetic field external to the disk, as suggested in Figure 3. The exact effect on the vectors on the edge of the disk depends on the field strength of the envelope and the  orientation of the field relative to the rotation axis\cite{Li2013}. The incoming fields may strongly influence or even dominate the magnetic field in the outer disk. In any case, the discrepancy indicates that, at least for this particular source, one needs to go beyond the simplest MRI-driven disk model that does not include any external influence or distortion. Both theoretical studies tailored to HL Tau and higher resolution polarization observations are needed to resolve this puzzle. 

If the disk of HL Tau has a dominant toroidal component, then it is uncertain what causes the outer vectors to not fit the toroidal morphology based on HL Tau's best fit disk dust model\cite{Kwon2011} (Figure 2, left). Perhaps the grains in some parts of the disk do not efficiently align with the disk field. Another possibility is that the magnetic field toward the inner disk is toroidal and beam-averaged to be aligned with the major axis of the disk. However, toward the edge of the disk, where the disk field may be less tightly wound and weaker, the magnetic field may be dominated by external field lines that are already toroidal (e.g., due to a rotating envelope). The incoming fields may strongly influence or even dominate the magnetic field in the outer disk. In any case, the discrepancy indicates that, at least for this particular source, one needs to go beyond the simplest magnetorotational instability\cite{Balbus1998} disk models that do not include any external influence or distortion. Both theoretical studies tailored to HL Tau and higher resolution polarization observations are needed to resolve this puzzle.

%Another possibility is that the external field lines entering the disk are already toroidal (e.g., due to a rotating envelope), and these incoming fields dominate the magnetic field at the edge of the disk. The more tightly-wound and denser inner part of the disk has a toroidal field that is beam-averaged to be aligned with the disk's major axis. Such a cartoon is seen in Figure \ref{hourglass} and could be responsible for the hint of an hourglass morphology seen in Figure 1.

%Another possibility is that the magnetic field toward the inner disk is toroidal and beam-averaged to be aligned with the major axis of the disk. However, toward the edge of the disk, where the disk field is less tightly wound and weaker, the magnetic field is dominated by external field lines that are already toroidal (e.g., due to a rotating envelope). Such a cartoon is seen in Figure \ref{hourglass} and could be responsible for the hint of an hourglass morphology seen in Figure 1.

%Regardless, observations do not conform to our simplest expectation, which indicates that the current theory is incomplete. Future work, both theoretical and observational, is needed to solve this interesting puzzle. 

At the 1000 AU scale, structured magnetic fields are observed around young, low-mass protostars\cite{Girart2006,Stephens2013}, but fields appear to be randomly aligned with respect to the inferred disk\cite{Hull2013,Hull2014}. Misalignment of the field lines with the rotation axis can help overcome magnetic braking to create a centrifugally supported disk at the 100 AU scale\cite{Hennebelle2009}. Further disk evolution can be driven by magnetorotational instability\cite{Balbus1998} or a magnetocentrifugal wind\cite{Blandford1982,Konigl2000}. A toroidally dominant disk field is expected in the former scenario, and a significant poloidal field is required for the latter (at least near the wind launching surface). Both processes can possibly contribute to the disk accretion at the same time. 

%As this protostellar disk forms, magnetorotational instability is thought to generally drive fields toward toroidal morphologies\cite{Balbus1998} while magnetocentrifugal winds drive fields toward poloidal morphologies\cite{Blandford1982,Konigl2000}. However, these processes can be less than ideal due to magnetic pressure, causing a mix of field components\cite{Hennebelle2009} .

%However, alignment can change between large-scale (several thousands of AU) and small-scale (several hundreds of AU), especially in weakly polarized sources\cite{Hull2014}. At the $\sim$100 AU scale, the formation of the disks can be suppressed by magnetic braking\cite{Mouschovias1991}, especially in cases where the initial magnetic field is aligned with the with the rotation axis\cite{Hennebelle2009}. Although it is somewhat uncertain how disks can form despite of magnetic braking\cite{Li2013}, magnetorotational instability is thought to generally drive fields toward toroidal morphologies\cite{Balbus1998} while magnetocentrifugal winds drive fields toward poloidal morphologies\cite{Blandford1982,Konigl2000}. However, these processes can be less than ideal due to magnetic pressure, causing a mix of components\cite{Hennebelle2009}.

Until now, we have been unable to observationally constrain the magnetic field morphology in disks. Along with the Class 0 source IRAS 16293--2422~B, the observations of HL Tau show that a toroidal field component may last from the low-mass protostar's initial formation to the T Tauri star stage -- approximately the first 10$^6$ years of a protostar's life\cite{Evans2009}. The apparent absence of vertical fields for these observations implies that magnetocentrifugal winds driven along large-scale poloidal magnetic fields\cite{Wardle1993} are probably not the dominant mechanism for redistributing the disk's angular momentum during the accretion process of a star at the probed 80 AU size-scale. However, the morphology detected in HL Tau cannot be fully explained by a simple mix of toroidal and vertical components either, requiring future observations at both large scale and small scale to truly understand the role of magnetic fields in the formation of solar systems like our own.

\bibliography{hltau_nature}

\begin{addendum}
\item [Supplementary Information] is available in the online version of the paper.
 \item We acknowledge Richard L. Plambeck and Charles L. H. Hull for their consultation during the data reduction process and Charles F. Gammie for helpful discussions. This research made use of APLpy, an open-source plotting package for Python hosted at http://aplpy.github.com. Illinois and Maryland were supported by NSF AST-1139950 and AST-1139998 respectively. Support for CARMA construction was derived from the states of California, Illinois, and Maryland, the James S. McDonnell Foundation, the Gordon and Betty Moore Foundation, the Kenneth T. and Eileen L. Norris Foundation, the University of Chicago, the Associates of the California Institute of Technology, and the National Science Foundation. Ongoing CARMA development and operations are supported by the National Science Foundation under a cooperative agreement (NSF AST 08-38226) and by the CARMA partner universities.
 %\item[Competing Interests] The authors declare that they have no
%competing financial interests.
% \item[Correspondence] Correspondence and requests for materials
%should be addressed to A.B.C.~(email: myaddress@nowhere.edu).
\item [Author Contributions] Data acquisition and reduction was performed by I. Stephens, L. Looney, and M. Fern\'{a}ndez-L\'{o}pez. Polarization modeling was performed by W. Kwon and fitted by I. Stephens. All authors analyzed and discussed the observations and manuscript.
\item[Author Information] Reprints and permissions information is available at www.nature.com/reprints. The authors declare no competing financial interests. Readers are welcome to comment on the online version of the paper. Correspondence and requests for materials should be addressed to I. Stephens. (ianws@bu.edu).
\end{addendum}

%\newpage
%\noindent {\bf 1. Detected magnetic field morphology of HL Tau at 0.6$\arcsec$ resolution.} The polarization vectors have been rotated by 90$^{\circ}$ to show the inferred magnetic field orientation. Red vectors are detections $>3\sigma_P$ while blue vectors are detections between 2$\sigma_P$ and 3$\sigma_P$. We also do not show vectors when the signal-to-noise for Stokes $I$ is below 2. The sizes of the vector are proportional to the fractional polarization, $P$. Stokes $I$ contours are shown for [--3, 3, 4, 6, 10, 20, 40, 60, 80, 100]~$\times \sigma_I$, where $\sigma_I=2.1$~mJy~beam$^{-1}$.  Color scale shows the polarized intensity with the color bar in Jy~beam$^{-1}$. The dashed line shows the major axis of PA = $136^\circ$\cite{Kwon2011}. The synthesized beam is shown on the bottom right and has a size of $0.65\arcsec \times 0.56\arcsec$ and a PA $=79.5^\circ$.
%\\
%\\
%\noindent {\bf 2. Observed magnetic field morphology compared with models.} The observed 3$\sigma$ and 5$\sigma$ detections are shown in red and orange, respectively, while modeled vectors are shown in yellow. Black contours, color scale, and beam size are the same as Figure 1.  a) 100\% toroidal field model with PA $=136^\circ$ and $i=40^\circ$. b) 100\% toroidal field model with PA $=151^\circ$ and $i=89^\circ$.

\newpage
\begin{figure} Ê
\begin{center}
\includegraphics[width=89mm]{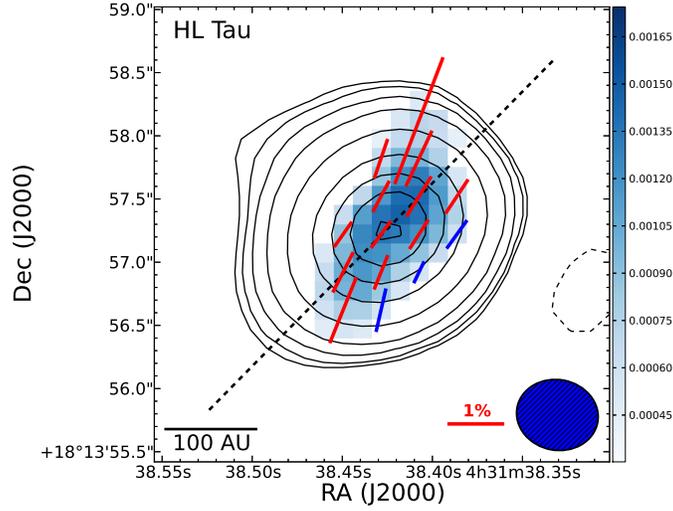}
\end{center}
\caption{ {\bf 1. Detected magnetic field morphology of HL Tau at 0.6$\arcsec$ resolution.} The polarization vectors have been rotated by 90$^{\circ}$ to show the inferred magnetic field orientation. Red vectors are detections $>3\sigma_P$ while blue vectors are detections between 2$\sigma_P$ and 3$\sigma_P$. We also do not show vectors when the signal-to-noise for Stokes $I$ is below 2. The sizes of the vector are proportional to the fractional polarization, $P$. Stokes $I$ contours are shown for [--3, 3, 4, 6, 10, 20, 40, 60, 80, 100]~$\times \sigma_I$, where $\sigma_I=2.1$~mJy~beam$^{-1}$.  Color scale shows the polarized intensity with the color bar in Jy~beam$^{-1}$. The dashed line shows the major axis of PA = $136^\circ$\cite{Kwon2011}. The synthesized beam is shown on the bottom right and has a size of $0.65\arcsec \times 0.56\arcsec$ and a PA $=79.5^\circ$.
}
\end{figure}

\begin{figure} 
\begin{center}
\hbox{\hspace{-5ex}\includegraphics[width=183mm]{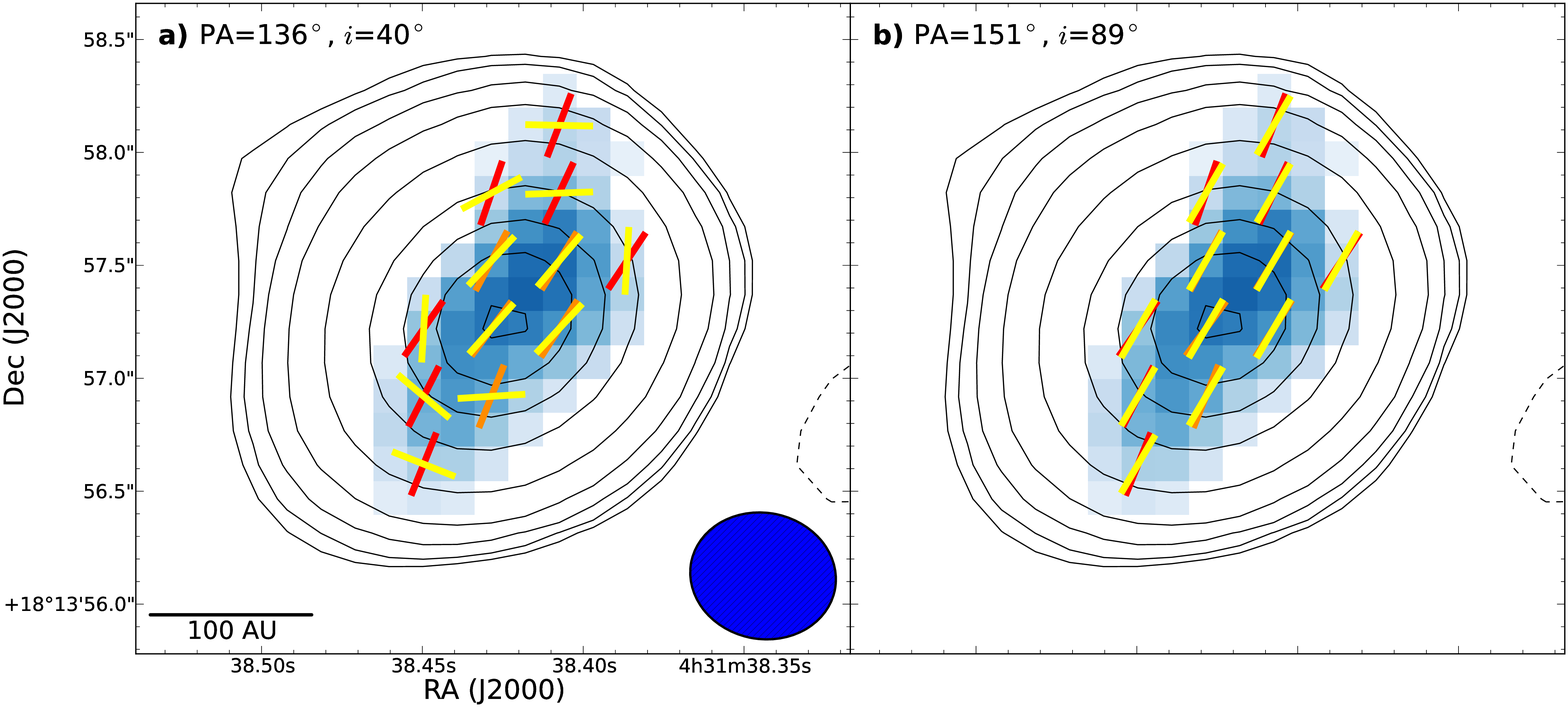}}
\caption{ {\bf 2. Observed magnetic field morphology compared with models.} The observed 3$\sigma$ and 5$\sigma$ detections are shown in red and orange, respectively, while modeled vectors are shown in yellow. Black contours, color scale, and beam size are the same as Figure 1.  a) 100\% toroidal field model with PA $=136^\circ$ and $i=40^\circ$. b) 100\% toroidal field model with PA $=151^\circ$ and $i=89^\circ$.
}
\end{center}
\end{figure}

\newpage
\begin{methods}
\subsection{Data Reduction.}

%CARMA continuum observations in Full Stokes mode were observed in two different resolutions with each resolution having multiple tracks. The C-array observations ($\sim$0.79$\arcsec$ resolution) consisted of 4 tracks in October and November 2013. After a successful detection of polarization, CARMA Director's Discretionary Time was awarded for B-array observations ($\sim$0.37$\arcsec$ resolution), and consisted of 3 more tracks in November and December 2013. The C-array observations were observed with a Local Oscillator (LO) frequency of 240 GHz, while the B-array observations used an LO frequency of 234 GHz. The latter used a different frequency since two of the CARMA antennas were unable to lock at 240 GHz.

%$^{27}$
%\cite{Sault1995}
%\cite{Witzel1986}
%\cite{Tafoya2004}
%\cite{MarroneRao2008}
%\cite{Dullemond2004}

The CARMA dual polarization receivers allow for the measurement of polarized dust emission. Polarimetric maps can thus be created, which provide the flux density (Stokes $I$), the PA of the dust polarization, and $P$ at every point within the map. CARMA continuum observations in Full Stokes mode were taken at 237 GHz in two different resolutions with each resolution having multiple tracks. The C-array observations ($\sim$0.79$\arcsec$ resolution) consisted of four tracks in October and November 2013 while the B-array observations ($\sim$0.37$\arcsec$ resolution) consisted of three more tracks in November and December 2013.

We have reduced the CARMA observations using the MIRIAD package$^{28}$. The dual-polarization receivers of CARMA measure right- (R) and left-circular (L) polarization and the four cross-polarization (RR, LL, LR, RL) terms. In order to calibrate CARMA Full Stokes observations, apart from the usual interferometric calibrations (bandpass, phase, and flux corrections), two additional calibrations are required: XYPhase (due to L and R channel phase differences) and leakage (due to L and R channels cross-coupling). These calibrations were done in the typical manner for CARMA\cite{Hull2014}. The leakage terms for each antenna were consistent from track to track, and the overall accuracy of the leakage calibrations are expected to be approximately 0.1\%. Day to day consistency in polarization observations was tested by measuring the phase calibrator, 0510+180. The polarization angle changed slightly from day to day, with a steady increase of a few degrees with each newer track. Intraday variability is a well-known phenomenon which affects the total flux density, the linearly polarized flux density, and the polarization angle$^{29}$ and likely explains the variations of a few degrees seen in 0510+180 from track to track. Since the variation of the polarization in 0510+180 was not very large, the consistency of polarization measurements of 0510+180 between tracks made us confident that our calibration is accurate. For B-array tracks, we also saw that the bandpass calibrator, 3C454.3, showed consistency for polarization measurements for all the tracks. Other calibrators observed did not have polarization detected, signifying that our polarization detection of HL Tau is not a spurious detection. We also note that there was a slight difference in the polarization angle between the lower and upper sideband; this difference may be due to Faraday rotation and was almost constant for all tracks.

For bandpass calibration, C-array observations used 3C84 and B-array observations used 3C454.3. The phase calibrator used for both arrays was 0510+180. Observations of MWC349, with an adopted flux of 2.1 Jy$^{30}$ provided the absolute scale for the flux calibration at 237 GHz in most of the tracks. The bootstrapped flux of 0510+180 using MWC349 was consistent within 10\% and 15\% to bootstrapped fluxes using Mars and Uranus respectively in other tracks. Since planets are resolved at these resolutions, MWC349 is likely to have a more accurate flux calibration and was bootstrapped for all tracks. The absolute flux uncertainty is estimated to be 15\%, but only statistical uncertainties are discussed in this work. When imaging, natural weighting was used to maximize the sensitivity.

Detections of polarization were found in every HL Tau track with consistent polarization angles and measurements. Since polarization is calculated from Stokes $Q$ and $U$ and can only have positive values, there exists a bias in the polarized intensity; hence, all our polarization measurements have been de-biased\cite{Hull2014}. The sensitivity in Stokes $I$ is limited by dynamic range rather than the flux sensitivity of the observations. The uncertainty in the absolute position angle of CARMA is approximately 3$^\circ$\cite{Hull2014}.

These high resolution interferometric observations are insensitive to large scale structure. Single dish and interferometric HL Tau observations find very similar compact fluxes, suggesting that the envelope dust continuum is negligible\cite{Lay1997}, and the flux appears to come entirely from the disk\cite{Kwon2011}. %Therefore, it is extremely likely that the polarization comes from the disk.

Also reported in this study are unpublished polarization observations (PA and $P$) of HL Tau from the SMA. These observations were taken in the compact configuration in October 2005 in the 345 GHz atmospheric window with the Local Oscillator tuned to 341.5 GHz. The polarization data reduction process was done in the typical manner employed by the SMA$^{31}$.

\subsection{Linear Polarization Modeling.}

We employed a flared viscous accretion disk model that was
constrained by high angular resolution
data and a broad spectral energy distribution, detailed in another paper\cite{Kwon2011}.  The accretion disk
model has a power-law density distribution with an exponential
tapering, and the vertical density distribution is assumed as
1.5 times thicker than the hydrostatic equilibrium case.
%$\rho(R,z)\prop(R/R_c)^{-p}\textrm{exp}(-(R/R_c)^{}.  $
The temperature distributions are calculated by interpolation of two power-law distributions at the cold mid-plane, comparable to the results of the Monte-Carlo
radiative code RADMC-3D$^{32}$ and at the
surface based on radiation equilibrium: $T_m[K]=190(r/\mbox{AU})^{-0.43}$ and $T_s[K]=600(r/\mbox{AU})^{-0.43}$.
The disk parameters employed were the volume density power-law index
$p=1.064$, the dust opacity spectral index $\beta=0.729$,
the disk mass $M_d=0.1349~M_\odot$, the inner radius $R_{in}=2.4$ AU,
and the characteristic radius $R_c=78.9$ AU. As described in the main
text, we investigated various cases of different inclination
and position angles. 

Our polarization modeling consists of toroidal and vertical magnetic
fields.  Instead of constraining the detailed morphology of magnetic
fields, we intended to constrain which morphology is preferred.  In order to achieve this goal, we examined 101 cases
spanning over relative polarization fractions of the two orthogonal
fields in steps of 1\% (i.e., 100\% toroidal, 99\% toroidal and 1\% vertical, 98\% toroidal and 2\% vertical, ...).  For constructing linear polarization information,
we built Stokes $I$, $Q$, and $U$ maps by numerically solving radiative transfer (necessary for a thick disk).  In individual integral
elements of radiative transfer along line of sight, we compute the
intensity for $Q$ and $U$. The fractional intensities
added up to the $Q$ and $U$ maps by an integral element are:
\begin{eqnarray}
\Delta Q = \Delta I ~ f_p (f_{tor} q_{tor} + f_{ver} q_{ver})\\
\Delta U = \Delta I ~ f_p (f_{tor} u_{tor} + f_{ver} u_{ver}),
\end{eqnarray}
where $f_p$ is a total polarization fraction ($\sqrt{Q^2+U^2}$/$I$),
$f_{tor}$ and $f_{ver}$ are relative fractions of the toroidal and
vertical fields (e.g., $f_{tor}=0.7$ and $f_{ver}=0.3$ for 70\% toroidal
and 30\% vertical fields), and $q_{tor}$
and $u_{tor}$ are cos$(2\chi_{tor})$ and sin$(2\chi_{tor})$
respectively. $\chi$ is the angle of the magnetic field measured
counterclockwise from the north. Similarly, $q_{ver}$ and $u_{ver}$ for
vertical fields are cos$(2\chi_{ver})$ and sin$(2\chi_{ver})$. Since the
disk is optically thin and we only care about the morphology, $f_p$ can
be given an arbitrary value (e.g., 0.01 or 0.1). Note that the toroidal and vertical magnetic field vectors at each
integral element have been tilted and rotated based on the inclination
and the position angle of the model disk, before the calculation
of the fractions. $Q$ and $U$ maps are convolved with the synthesized beam from
the polarization observations, and the modeled position angles of the magnetic field morphology is created
using $\chi=0.5~\textrm{tan}^{-1}$($U$/$Q$).
\end{methods}

%\newpage
%\begin{figure} Ê
%
%\includegraphics[width=89mm]{HL_Tau_aplpy.eps}
%
%\end{figure}
%
%\begin{figure} Ê
%\includegraphics[width=183mm]{HL_Tau_2models.eps}
%\end{figure}
%
%
%
%
%\begin{figure} Ê
%\includegraphics[scale=0.5]{toroid_outer.png}
%\end{figure}

\end{document}